\documentclass{aa}  
\usepackage{graphicx}
\usepackage{txfonts}
\begin{document}
   \title{Angular Correlation of the CMB in the $R_{\rm h}=ct$ Universe}

   \author{F. Melia
\thanks{John Woodruff Simpson Fellow}
          }

   \offprints{F. Melia}

   \institute{Department of Physics, The Applied Math Program, and Department of Astronomy, 
The University of Arizona, Tucson, Arizona 85721, USA \\
\email{fmelia@email.arizona.edu}
          }

   \date{Received July 15, 2013}

 
  \abstract
   {The emergence of several unexpected large-scale features in the cosmic microwave background
(CMB) has pointed to possible new physics driving the origin of density fluctuations in the early 
Universe and their evolution into the large-scale structure we see today.
 } 
   {In this paper, we focus our attention on the possible absence of angular correlation in 
the CMB anisotropies at angles larger than $\sim 60^\circ$, and consider whether this feature 
may be the signature of fluctuations expected in the $R_{\rm h}=ct$ Universe.
}
   {We calculate the CMB angular correlation function for a fluctuation spectrum
expected from growth in a Universe whose dynamics is constrained by the 
equation-of-state $p=-\rho/3$, where $p$ and $\rho$ are the total pressure and
density, respectively. 
}
   {We find that, though the disparity between the predictions of $\Lambda$CDM 
and the WMAP sky may be due to cosmic variance, it may also be due to an absence of 
inflation. The classic horizon problem does not exist in the $R_{\rm h}=ct$
Universe, so a period of exponential growth was not necessary in this
cosmology in order to account for the general uniformity of the CMB (save for
the aforementioned tiny fluctuations of 1 part in 100,000 in the WMAP relic
signal). 
}
   {We show that the $R_{\rm h}=ct$ Universe without inflation can
account for the apparent absence in CMB angular correlation at angles
$\theta\ga 60^\circ$ without invoking cosmic variance,
providing additional motivation for pursuing this cosmology
as a viable description of nature.
}

   \keywords{cosmic microwave background -- cosmological parameters -- cosmology:
    observations -- cosmology: theory -- cosmology: dark matter -- gravitation
               }

   \maketitle
%

\section{Introduction}
The high signal-to-noise maps of the cosmic microwave background (CMB)
anisotropies, particularly those produced by the Wilkinson Microwave Anisotropy
Probe (WMAP; Bennett et al. 2003; Spergel et al. 2003) and, more recently,
by {\it Planck} (Planck Collaboration XV 2013), have revolutionized our ability 
to probe the Universe on its largest scales. In the near future,
even higher resolution temperature maps and high-resolution polarization maps,
perhaps also tomographic 21-cm observations, will extend our knowledge of
the Universe's spacetime and its fluctuations to a deeper level, possibly
probing beyond the surface of last scattering.

Yet the emergence of greater detail in these all-sky maps has revealed 
several possible unexpected features on large scales, some of which 
were first reported by the Cosmic Background Explorer (COBE) Differential 
Microwave Radiometer (DMR) collaboration (Wright et al. 1996). These include 
an apparent alignment of the largest modes of CMB anisotropy, as well
as unusually low angular correlations on the largest scales.

Though viewed as significant anomalies at first (Spergel et al. 2003), 
these unexpected features may now be explained as possibly being due to cosmic
variance within the standard model (Bennett et al. 2013). However, they may
also be interesting and important for several reasons. Chief among them is the
widely held view that the large-scale structure in the present Universe developed
via the process of gravitational instability from tiny primordial fluctuations in
energy density. The temperature fluctuations in the CMB, emerging several
hundred thousand years after the big bang, are thought to be associated with
the high-redshift precursors of the fluctuations that generated the galaxies
and clusters we see today. Therefore, an absence of correlations
in the CMB anisotropies may hint at required modifications to the standard model
($\Lambda$CDM), or possibly even new physics, each of which may alter
our view of how the Universe came into existence and how it evolved from
its earliest moments to the present state.

Our focus in this paper will be the possible absence of angular correlation
in the CMB at angles larger than $\sim 60^\circ$. This feature may be anomalous
because the absence of any angular correlation at the largest scales would
be at odds with the inflationary paradigm (Guth 1981; Linde 1982). But 
without inflation, $\Lambda$CDM simply could not account for the apparent 
uniformity of the CMB (other than fluctuations at the level of 1 part in 
100,000 seen in the WMAP relic signal) across the sky. Thus, if variance
is not the cause of the apparent disparity, the standard model of cosmology 
would be caught between contradictory observational constraints.

In this paper, we will therefore explore the possibility that these
possible CMB anomalies might be understood within the context of the recently
introduced $R_{\rm h}=ct$ Universe. This cosmology is motivated by a
strict adherence to the requirements of the Cosmological Principle
and the Weyl postulate, which together suggest that the Universe
must be expanding at a constant rate. Additional theoretical support
for this conclusion was reached with the recent demonstration that
the Friedmann-Robertson-Walker (FRW) metric is apparently only valid for
a perfect fluid with zero active mass, i.e., with $\rho+3p=0$, in terms
of the total energy density $\rho$ and pressure $p$ (Melia 2013b);
this is the equation-of-state that gives rise to the $R_{\rm h}=ct$
condition. We recently showed that
the horizon problem, so evident in $\Lambda$CDM, actually does not exist
in the $R_{\rm h}=ct$ Universe (Melia 2013a), so inflation is not
required to bring the CMB into thermal equilibrium following the big bang.
The $R_{\rm h}=ct$ Universe without inflation should therefore provide
a  meaningful alternative to $\Lambda$CDM for the purpose of interpreting
the CMB angular correlations.

\section{The $R_{\rm h}=ct$ Universe}
The $R_{\rm h}=ct$ cosmology is still at an early stage of development and,
given that its origin and structure may not yet be well known, we will begin
by describing its principal features. There are several ways of looking at
the expansion of the Universe. One is to guess its constituents and
their equation of state and then solve the dynamics equations to determine
the expansion rate as a function of time. This is the approach taken by
$\Lambda$CDM. The second is to use symmetry arguments and our knowledge
of the properties of a gravitational horizon in general relativity (GR)
to determine the spacetime curvature, and thereby the expansion rate,
strictly from just the value of the total energy density $\rho$ and the
implied geometry, without necessarily having to worry about the specifics
of the constituents that make up the density itself. This is the approach
adopted by $R_{\rm h}=ct$. The constituents of the Universe must then
partition themselves in such a way as to satisfy that expansion rate.
In other words, what matters is $\rho$ and the overall equation of
state $p=w\rho$, in terms of the total pressure $p$ and total
energy density $\rho$. In $\Lambda$CDM, one assumes $\rho=\rho_m+
\rho_r+\rho_{de}$, i.e., that the principal constituents are matter,
radiation, and an unknown dark energy, and then infers $w$ from the
equations of state assigned to each of these constituents. In
$R_{\rm h}=ct$, it is the aforementioned symmetries and other
constraints from GR that uniquely fix $w$.

The $R_{\rm h}=ct$ Universe is an FRW
cosmology in which Weyl's postulate takes on a more important role
than has been considered before (Melia \& Shevchuk 2012). There
is no modification to GR,
and the Cosmological principle is adopted from the start, just
like any other FRW cosmology. However, Weyl's postulate adds a very
important ingredient. Most workers assume that Weyl's postulate
is already incorporated into all FRW metrics, but actually it is only
partially incorporated. Simply stated, Weyl's postulate says that any
proper distance $R(t)$ must be the product of a universal expansion
factor $a(t)$ and an unchanging co-moving radius $r$, such that
$R(t)=a(t)r$. The conventional way of writing an FRW metric adopts this
coordinate definition, along with the cosmic time $t$, which is actually
the observer's proper time at his/her location. But what is often
overlooked is the fact that the gravitational radius, $R_{\rm h}$,
which has the same definition as the Schwarzschild radius, and
actually coincides with the better known Hubble radius, is in
fact itself a proper distance too (see also Melia and Abdelqader 2009).
And when one forces this radius
to comply with Weyl's postulate, there is only one possible choice for
$a(t)$, i.e., $a(t)=(t/t_0)$, where $t_0$ is the current age of the
Universe. This also leads to the result that the gravitational radius must
be receding from us at speed $c$, which is in fact how the Hubble radius
was defined in the first place, even before it was recognized as another
manifestation of the gravitational horizon. Those familiar with black-hole
spacetimes already know that a free-falling observer sees the event horizon
approaching them at speed $c$, so this property of $R_{\rm h}=ct$ is not
surprising in the context of GR.

The principal difference between $\Lambda$CDM and $R_{\rm h}=ct$ is
how they handle $\rho$ and $p$. In the $R_{\rm h}=ct$ cosmology, the fact that
$a(t)\propto t$ requires that the total pressure $p$ be given as
$p=-\rho/3$ (and, as we have already noted, it now appears that the
FRW metric is only valid when the active mass $\rho+3p$ is exactly zero). 
The consequence of this is that quantities such as the
luminosity distance and the redshift dependence of the Hubble constant
$H$, take on very simple, analytical forms. Though we won't necessarily
need to use these here, we quote them for reference. In $R_{\rm h}=ct$,
the luminosity distance is
\begin{equation}
d_L=R_h(t_0)(1+z)\ln(1+z)\;,
\end{equation}
and
\begin{equation}
H(z)=H_0(1+z)\;,
\end{equation}
where $z$ is the redshift, $R_h=c/H$, and $H_0$ is the value of the
Hubble constant today. These relations are clearly very relevant to
a proper examination of other cosmological observations, and we are
in the process of applying them accordingly. For example, we have
recently demonstrated that the model-independent cosmic chronometer
data (see, e.g., Moresco et al. 2012) are a better match to
$R_{\rm h}=ct$ (using Eq. 2), than the concordance, best-fit
$\Lambda$CDM model (see Melia \& Maier 2013). The same applies
to the gamma-ray burst Hubble diagram (Wei et al. 2013).

In the end, regardless of how $\Lambda$CDM or $R_{\rm h}=ct$ handle
$\rho$ and $p$, they must both account for the same cosmological
data. There is growing evidence that, with its empirical approach,
$\Lambda$CDM can function as a reasonable approximation to
$R_{\rm h}=ct$ in some restricted redshift ranges, but apparently
does poorly in others (such as the topic under consideration in this
paper). For example, in using the ansatz $\rho=\rho_m+\rho_r+\rho_{de}$
to fit the data, one finds that the $\Lambda$CDM parameters must have
quite specific values, such as $\Omega_m\equiv \rho_m/\rho_c=0.27$
and $w_{de}=-1$, where $\rho_c$ is the critical density and $w_{de}$
is the equation-of-state parameter for dark energy. This is quite
telling because with these parameters, $\Lambda$CDM then requires
$R_{\rm h}(t_0)=ct_0$ today. That is, the best-fit $\Lambda$CDM
parameters describe a universal expansion equal to what it would
have been with $R_{\rm h}=ct$ all along. Other indicators support
the view that using $\Lambda$CDM to fit the data therefore
produces a cosmology almost (but not entirely) identical to
$R_{\rm h}=ct$ (see Melia 2013c).

For example, by allowing each of its constituents (matter, radiation,
and dark energy) to vary according to their assumed dependencies on
$a(t)$, without the global restriction that $R_{\rm h}$ must be equal
to $ct$ for all $t$, the value of $R_{\rm h}$ predicted by $\Lambda$CDM
fluctuates about the mean it would otherwise always have if the constraint
$R_{\rm h}=ct$ were imposed from the start. So $\Lambda$CDM finds itself
in this awkward situation in which the value of $R_{\rm h}(t_0)$ today is
forced to equal $ct_0$, but in order to achieve this ``coincidence," the
Universe had to decelerate early on, followed by a more recent acceleration
that exactly balanced out the effects of its slowing down at the beginning.
As shown in (Melia 2013a), it is specifically this early deceleration in
$\Lambda$CDM that brings it into conflict with the near uniformity of the
CMB data, requiring the introduction of an inflationary phase to rescue it.

This important difference between $\Lambda$CDM and $R_{\rm h}=ct$ means
that fluctuation growth was driven to all scales in the former cosmology,
meaning that we should now see an angular correlation at all angles
across the sky. However, since $R_{\rm h}=ct$ was not subject to this
early exponential growth, its fluctuations in the CMB were limited
in size by the gravitational horizon at the time of recombination.
We will show below that this limit results in an absence of angular
correlation at angles greater than about $60^\circ$, which is what
the data seem to suggest. This property of $R_{\rm h}=ct$ also correctly
accounts for the location, $\theta_{\rm min}$, of the minimum in
the angular correlation function $C(\cos\theta)$ and its value,
$C(\cos\theta_{\rm min})$, at that angle.

\section{The Angular Correlation Function of the CMB}
Assuming that the statistical distributions of matter and metric fluctuations about
the background metric are isotropic, the CMB temperature seen in direction
$\hat{\bf n}$ is predicted to be described by a Gaussian random field on the
sky, implying that we can expand it in terms of spherical harmonics $Y_{lm}(
\hat{\bf n})$, using independent Gaussian random coefficients $a_{lm}$ of
zero mean. The two-point correlation (for directions $\hat{\bf n}_1$ and
$\hat{\bf n}_2$) becomes a function of $\cos\theta\equiv \hat{\bf n}_1\cdot
\hat{\bf n}_2$ only and can be expanded in terms of Legendre polynomials:
\begin{equation}
C(\cos\theta)\equiv \langle T(\hat{\bf n}_1)T(\hat{\bf n}_2)\rangle=
{1\over 4\pi}\sum_l (2l+1)C_l P_l(\cos\theta)\;.
\end{equation}
Statistical independence implies that
\begin{equation}
\langle a^*_{lm}a_{l'm'}\rangle\propto\delta_{ll'}\,\delta{mm'}\;,
\end{equation}
and statistical isotropy further requires that the constant of proportionality
depend only on $l$, not $m$:
\begin{equation}
\langle a^*_{lm}a_{l'm'}\rangle=C_l\,\delta_{ll'}\,\delta{mm'}\;.
\end{equation}
The constant
\begin{equation}
C_l={1\over 2l+1}\sum_m |a_{lm}|^2
\end{equation}
is the angular power of the multipole $l$.

To properly calculate the CMB angular correlation function, one must first navigate through
a complex array of processes generating the incipient fluctuations $\Delta\rho$ in density, followed by a
comparably daunting collection of astrophysical effects, all subsumed together into multiplicative
factors known as ``transfer functions," that relate the power spectrum $P(\kappa)$ of the
gravitational potential resulting from $\Delta\rho$ to the output CMB temperature
fluctuations $\Delta T$. A detailed calculation of this kind can take weeks or even
months to complete, even on advanced computer platforms.

Fluctuations produced prior to decoupling lead to ``primary" anisotropies,
whereas those developing as the CMB propagates from the surface of last scattering to
the observer are ``secondary."  The former include temperature variations associated with
photon propagation through fluctuations of the metric. Known as the Sachs-Wolfe effect
(Sachs \& Wolfe 1967), this process produces fluctuations in temperature given
roughly as
\begin{equation}
{\Delta T\over T}\approx {1\over 3}{\Delta\Phi\over c^2}\;,
\end{equation}
where $\Delta\Phi$ is a fluctuation in the gravitational potential.
The Sachs-Wolfe effect is dominant on large scales (i.e., $\theta\gg 1^\circ$).

Prior to decoupling, the plasma is also susceptible to acoustic oscillations. The
density variations associated with compression and rarefaction produce baryon fluctuations
resulting in a prominent acoustic peak seen at $l\sim 200$ in the power spectrum.
Processes such as this, which depend sensitively on the microphysics, are therefore
dominant on small scales, typically $\theta<1^\circ$.

Once the photons decouple from the baryons, the CMB must propagate through a
large scale structure with complex distributions of the gravitational potential and
intra-cluster gas, neither of which is necessarily isotropic or homogeneous on small
spatial scales. Astrophysical processes following recombination therefore imprint
their own (secondary) signatures on the CMB temperature anisotropies. Examples
of secondary processes include: the thermal and kinetic Sunyaev-Zeldovich effects
(Sunyaev \& Zeldovich 1980), due to inverse Compton scattering by, respectively,
thermal electrons in clusters and electrons moving in bulk with their galaxies
relative to the CMB ``rest" frame; the integrated Sachs-Wolfe effect, induced by
the time variation of gravitational potentials, and its non-linear extension, the
Rees-Sciama effect (Rees \& Sciama 1968); and the deflection of CMB light by
gravitational lensing. These effects are expected to produce observable temperature
fluctuations at a level of order $\Delta T/T\sim 10^{-5}$ on arc-minute scales (though
the Rees-Sciama effect typically generates much smaller fluctuations with $\Delta T/T\sim
10^{-8}$).

Insofar as understanding the global dynamics of the Universe is concerned, there are
several reasons why we can find good value in utilizing the angular correlation function
$C(\cos\theta)$. This is not to say that the power spectrum (represented by the
set of $C_l$'s) is itself not probative. On the contrary, the past forty years have shown
that a meaningful comparison can now be made between theory and observations
through an evaluation, or determination, of the multipole powers. But previous
studies have also shown that variations caused by different cosmological
parameters are not orthogonal, in the sense that somewhat similar sets of
$C_l$'s can be found for different parameter choices (Scott et al. 1995).

A principal reason for this is that the two-point angular power spectrum emphasizes
small scales (typically $\sim 1^\circ$), making it a useful diagnostic for physics at the
last scattering surface. A comparison between theory and observation on these small
scales permits the precise determination of fundamental cosmological parameters,
given an assumed cosmological model (Nolta et al. 2009). The two-point angular
correlation function $C(\cos\theta)$ contains the same information as the angular
power spectrum, but highlights behavior at large angles (i.e., small values of $l$), the
opposite of the two-point angular power spectrum. Therefore, the angular correlation
function provides a better test of dynamical models driving the universal expansion.

For these reasons, we suggest that a comparison between the calculated $C(\cos\theta)$
and the observations may provide a more stringent test of the assumed cosmology.
The microphysical effects responsible for the high-$l$ multipoles, featured most prominently
in the power spectrum, are more likely to be generic to a broad range of expansion
scenarios. But $C(\cos\theta)$, which highlights the largest scale fluctuations,
yields greater differentiation when it comes to the overall dynamics. A
more complete discussion of the benefits of the angular power spectrum
versus the angular correlation function, and vice versa, is provided in
Copi et al. (2013), and many other references cited therein.

\section{Angular Correlation Function of the CMB in $\Lambda$CDM}
Let us now consider the predicted function $C(\cos\theta)$ in the context of
$\Lambda$CDM and its comparison to the WMAP sky. A crucial ingredient of the
standard model is cosmological inflation---a brief phase of very rapid expansion
from approximately $10^{-35}$ seconds to $10^{-32}$ seconds following the
big bang, forcing the universe to expand much more rapidly than would otherwise
have been feasible solely under the influence of matter, radiation, and dark energy,
carrying causally connected regions beyond the horizon each would have had in the
absence of this temporary acceleration. In $\Lambda$CDM without this exponential
expansion, regions on opposite sides of the sky would not have had sufficient time
to equilibrate before producing the CMB at $t_e\sim 380,000$ years after the
big bang (see Melia 2013a for a detailed explanation and a comparison between
various FRW cosmologies). Therefore, the predictions
of $\Lambda$CDM would be in direct conflict with the observed uniformity of the
microwave background radiation, which has the same temperature everywhere,
save for the aforementioned fluctuations at the level of one part in 100,000 seen
in WMAP's measured relic signal.

This required inflationary expansion drives the growth of fluctuations 
on all scales and predicts an angular correlation at all angles. However,
as pointed out by Copi et al. (2013) in their detailed analysis of
both the WMAP and {\it Planck} data, there has always been an
indication (even from the older {\it COBE-DMR} observations; Hinshaw
et al. 1996) that the two-point angular correlation function nearly vanishes 
on scales greater than about 60 degrees, contrary to what 
$\Lambda$CDM predicts (see figure~1). From this figure, one may
also come away with the impression that there are significant
differences between the predicted and observed angular
correlation function at angles smaller than 60 degree, but because
the different angular bins are correlated, the deviation between
the two curves is not as statistically significant as it appears. In
fact, cosmic variance from the theoretical curve (indicated by the 
shaded region) can essentially account for most of the disparity
at these smaller angles.

   \begin{figure}[hp]
   \centering
      \includegraphics[angle=0,width=8.1cm]{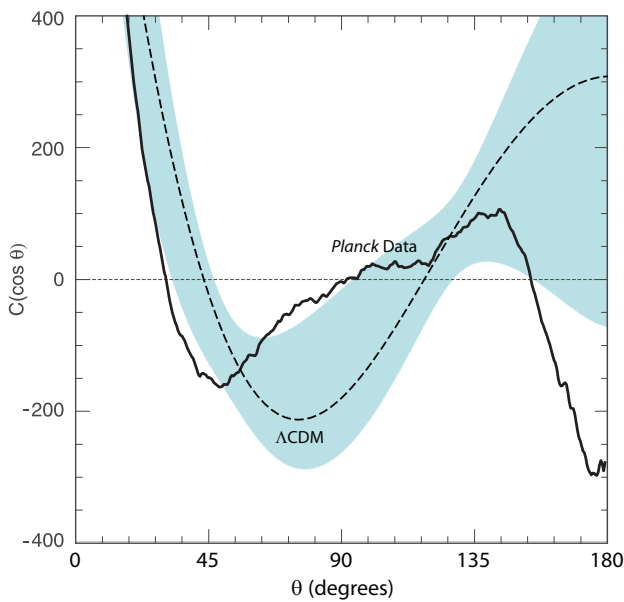}
      \caption{
Angular correlation function of the best-fit $\Lambda$CDM model and
that inferred from the {\it Planck} SMICA full-sky map (Planck Collaboration 
XV 2013) on large angular scales. The shaded region is the one-sigma 
cosmic variance interval. (This figure is adapted from Copi et al. 2013)}
\end{figure}

A quantitative measure of the differences between the 
observed angular correlation function and that predicted by 
$\Lambda$CDM is the so-called $S_{1/2}$ statistic 
introduced by the WMAP team (Spergel et al. 2003):
\begin{equation}
S_{1/2}=\int_1^{1/2}\left|C(\cos\theta)\right|^2d(\cos\theta)\;.
\end{equation}
A subsequent application of this measure on the WMAP 5 year maps (Copi 
et al. 2009) revealed that only $\sim 0.03\%$ of $\Lambda$CDM model CMB skies 
have lower values of $S_{1/2}$ than that of the observed WMAP sky.
But we may simply be dealing with foreground subtraction issues. The 
final (9-year) analysis of the WMAP data suggests that the difference 
between theory and observation is probably smaller and may fall within 
the $1-\sigma$ error region associated with cosmic variance (Bennett 
et al. 2013). 

There are indications, however, that the differences between theory
and observations may be due to more than just randomness. For example,
the observed angular correlation function has a well defined shape,
with a minimum at $\sim 50^\circ$, and a relatively smooth curve
on either side of this turning point. One might have expected the
data points to not line up as they do within the variance window
if stochastic processes were solely to blame. Moreover, one cannot
ignore the fact that the observed angular correlation goes to zero
beyond $\sim 60^\circ$. Variance could have resulted in a function
with a different slope than that predicted by $\Lambda$CDM, but
it seems unlikely that this randomly generated slope would be
close to zero above $\sim 60^\circ$.

The recent {\it Planck} results also confirm that the $S_{1/2}$
statistic is very low. In their analysis, Copi et al. (2013) find
that the probability of the observed cut-sky $S_{1/2}$ statistic in 
an ensemble of realizations of the best-fitting $\Lambda$CDM model 
never exceeds $0.33\%$ for any of the analyzed combinations
of maps and masks. This trend has remained intact since the release
of the WMAP 3-year data. The apparent lack of temperature
correlations on large angular scales is robust and increases
in statistical significance as the quality of the instrumentation
improves, suggesting that instrumental issues are not the
cause.

If it turns out that the absence of large-angle correlation is
real, and not due to cosmic variance, this may be the most 
significant result of the WMAP mission, because it essentially 
invalidates any role that inflation might have played in the 
universal expansion. Our principal goal in this paper is 
therefore to examine whether the $R_{\rm h}=ct$
Universe---a cosmology without inflation---can account for the
lack of temperature correlations on large angular scales
without invoking cosmic variance.

\section{Angular Correlation of the CMB in the $R_{\rm h}=ct$ Universe}
Since the $R_{\rm h}=ct$ Universe did not undergo a period of inflated expansion, 
it is not subject to the observational restriction discussed above. 
Defining the density contrast $\delta\equiv \delta\rho/\rho$ in terms
of the density fluctuation $\delta \rho$ and unperturbed density $\rho$,
we can form the wavelike decomposition
\begin{equation}
\delta=\sum_\kappa \delta_\kappa(t)e^{i\vec{\kappa}\cdot{\bf r}}\;,
\end{equation}
where the Fourier component $\delta_\kappa$ depends only on cosmic time
$t$, and $\vec{\kappa}$ and {\bf r} are the co-moving wavevector and radius,
respectively. In Melia \& Shevchuk (2012), we derived the dynamical equation for
$\delta_\kappa$ in the $R_{\rm h}=ct$ Universe and showed that in the linear
regime
\begin{equation}
\ddot{\delta}_\kappa+{3\over t}\dot{\delta}_\kappa={1\over 3}c^2\left({\kappa\over a}\right)^2\delta_\kappa\;.
\end{equation}

The way perturbation growth is handled in $R_{\rm h}=ct$, leading to Equation~(10),
 is somewhat different from $\Lambda$CDM, so we will take a moment to briefly 
describe the origin of this expression. As we discussed in the introduction,
the chief difference between $\Lambda$CDM and $R_{\rm h}=ct$ is that one 
must guess the constituents of $\rho$ in the former, assign an individual 
equation of state to each, and then solve the growth equation derived for 
each of these components separately. This is how one handles a situation
in which the various species do not necessarily feel each other's pressure, 
though they do feel the gravitational influence from the total density.
The coupled equations of growth for the various components can be quite 
complex, so one typically approximates the equations by expressions that
highlight the dominant species in any given era. For example, before 
recombination, the baryon and photon components must be treated 
as a single fluid, since they are coupled by frequent interactions in 
an optically-thick environment. During this period, $\Lambda$CDM
assumes that ``dark energy" is smooth on scales corresponding 
to the fluctuation growth, and treats the baryon-photon fluid 
as a single perturbed entity with the pressure of radiation and 
an overall energy density corresponding to their sum. 
Once the radiation decouples from the luminous
matter, all four constituents (including dark matter) must be 
handled separately, though in a simplified approach one may 
ignore the radiation, which becomes sub-dominant at later times.

The situation in $R_{\rm h}=ct$ is quite different for several 
reasons. First of all, the overall equation of state in this cosmology 
is not forced on the system by the constituents; it is the other way 
around. The symmetries implied by the Cosmological Principle and 
Weyl's postulate together, through the application of general
relativity, only permit a constant expansion rate, which means 
that $p = -\rho/3$.  The expansion rate depends on the total 
energy density, but not on the partitioning among the various 
constituents. Instead, the constituents must partition themselves 
in such a way as to always guarantee that this overall equation
of state is maintained during the expansion.

   \begin{figure}[hp]
   \centering
      \includegraphics[angle=0,width=8.2cm]{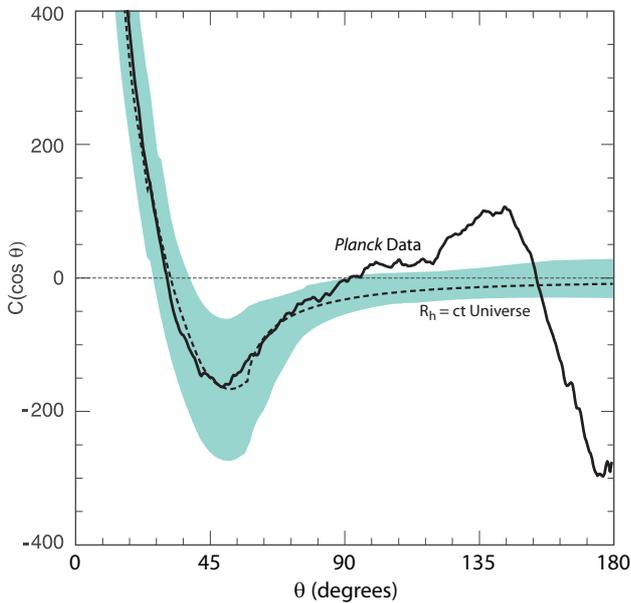}
      \caption{
Angular correlation function of the CMB in the $R_{\rm h}=ct$ Universe,
for $b=3$ and $t_0/t_e=5\times 10^5$ (see text). The {\it Planck}
data are the same as those shown in figure~1. The shaded region
is the one-sigma cosmic variance interval.}
\end{figure}

And since the pressure is therefore a non-negligible fraction of $\rho$
at all times, one cannot use the equations of growth derived
from Newtonian theory (commonly employed in $\Lambda$CDM), 
since $p$ itself acts a source of curvature. One must therefore 
necessarily start with the relativistic growth equation (numbered 
41 in Melia \& Shevchuk 2012), which correctly incorporates all of the 
contributions from $\rho$ and $p$. This equation is ultimately derived 
from Einstein's field equations using the perfect-fluid form of the 
stress-energy tensor, written in terms of the total $\rho$ and total 
$p$, but without specifying the sub-partitioning of the density
and pressure among the various constituents. With this approach,
there is only one growth equation.

On occasion, it is also necessary to use the relativistic growth 
equation in $\Lambda$CDM. But there, one typically chooses a 
regime where a single constituent is dominant, say during the 
matter-dominated era, and then one assumes that $\rho$ is 
essentially just the density due to matter (for which also
$p\approx 0$). But in general, since the pressure appearing 
in the stress-energy tensor is the total pressure, one cannot 
mix and match different components that may or may not ``feel" 
each other's influence (as described above). So in fact using 
the relativistically correct growth equation is difficult in $\Lambda$CDM, 
unless one can make suitable approximations in a given regime. 

In $R_{\rm h}=ct$, on the other hand, the total pressure is 
always $-\rho/3$, so the key question is whether all of the
constituents participate in the perturbation growth, or 
whether only some of them do. There is no doubt that the 
baryons and photons are coupled prior to recombination. In 
$\Lambda$CDM, one assumes that dark energy is coupled
only weakly, acting as a smooth background. In $R_{\rm h}=ct$, dark energy 
cannot be a cosmological constant. One therefore assumes that 
during the early fluctuation growth, everything is coupled strongly
in order to maintain the required total pressure $-\rho/3$. 

In short, there is one assumption made in each cosmology. In $\Lambda$CDM, 
dark energy is a cosmological constant that remains smooth while the
the baryon-photon fluid is perturbed at early times. In $R_{\rm h}=ct$, 
dark energy cannot be a cosmological constant, and everything is 
coupled strongly at early times, so the perturbation affects the
total energy density $\rho$. One must always use the correct 
relativistic growth equation, which includes $p$ as a source
of gravity.

In the end, this equation simplifies considerably because the active mass 
in $R_{\rm h}=ct$  is proportional to $\rho+3p=0$, and therefore the gravitational term
normally appearing in the standard model is absent. But this does not mean that $\delta_\kappa$
cannot grow. Instead, because $p<0$, the (usually dissipative) pressure term on the right-hand-side
here becomes an agent of growth. Moreover, there is no Jeans length scale. In its place is the
gravitational radius, which we can see most easily by recasting this differential equation in the form
\begin{equation}
\ddot{\delta}_\kappa+{3\over t}\dot{\delta}_\kappa-{1\over 3}{\Delta_\kappa^2\over t^2}\delta_\kappa=0\;,
\end{equation}
where
\begin{equation}
\Delta_\kappa\equiv {2\pi R_{\rm h}\over \lambda}\;.
\end{equation}
Note, in  particular, that both  the gravitational radius $R_{\rm h}$ and the fluctuation
scale $\lambda$ vary with $t$ in exactly the same way, so $\Delta_\kappa$ is therefore
a constant in time. But the growth rate of $\delta_\kappa$ depends critically on whether
$\lambda$ is less than or greater than $2\pi R_{\rm h}$.

The fact that $R_{\rm h}=ct$ does not have a Jeans length is itself quite
relevant to understanding the observed lack of any scale dependence in the 
measured matter correlation function (Watson et al. 2011). As one can see
from the general form of the dynamical equation for $\delta_\kappa$ (Melia
\& Shevchuk 2012), only a cosmology with $p=-\rho/3$ has this feature. In
every other case, both the pressure and gravitational terms are present
in Eq.~(10), which always produces a Jeans scale. For example, $\Lambda$CDM
predicts different functional forms for the matter correlation function
on different spatial scales, and is therefore not consistent with the
observed matter distribution.

A simple solution to Eq.~(11) is the power law
\begin{equation}
\delta_\kappa(t)=\delta_\kappa(0)t^\alpha\;,
\end{equation}
where evidently
\begin{equation}
\alpha^2+2\alpha-{1\over 3}\Delta^2_\kappa=0\;,
\end{equation}
so that
\begin{equation}
\alpha=-1\pm\sqrt{1+\Delta_\kappa^2/3}\;.
\end{equation}
Thus, for small fluctuations ($\lambda<<2\pi R_{\rm h}$), the growing mode is
\begin{equation}
\delta_\kappa\sim \delta_\kappa(0)t^{\Delta_\kappa/\sqrt{3}}\;,
\end{equation}
whereas for large fluctuations ($\lambda>2\pi R_{\rm h}$), the dominant mode
\begin{equation}
\delta_\kappa\sim \delta_\kappa(0)
\end{equation}
does not even grow. For both small and large fluctuations, the second mode decays away.

   \begin{figure}[hp]
   \centering
      \includegraphics[angle=0,width=8.2cm]{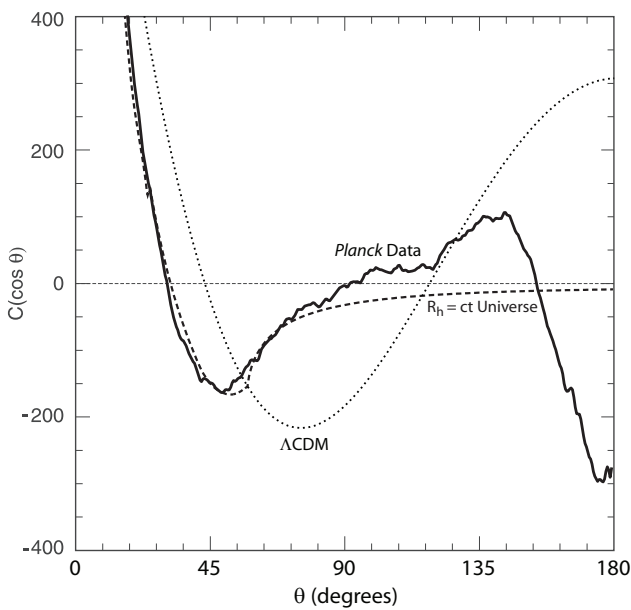}
      \caption{Angular correlation function of the CMB in the $R_{\rm h}=ct$ Universe,
      for $b=3$ and $t_0/t_e=5\times 10^5$, together with the best-fit $\Lambda$CDM model,
      compared with the {\it Planck} data.}
\end{figure}

A quick inspection of the growth rate implied by Eq.~(16) reveals that the fluctuation
spectrum must have entered the non-linear regime well before recombination
(which presumably occurred at $t_e\sim 10^4$--$10^5$ years). Therefore, without carrying out
a detailed simulation of the fluctuation growth at early times, it is not possible to extract the
power spectrum
\begin{equation}
P(\kappa)=\langle |\delta_\kappa|^2\rangle
\end{equation}
unambiguously. Fortunately, this is not the key physical ingredient we are seeking. Instead,
the central question is ``What is the maximum range over which the fluctuations would have
grown, either linearly or non-linearly, prior to recombination?" and this we can answer rather
straightforwardly.

Eqs.~(16) and (17) suggest that---without inflation---the maximum fluctuation size
at any given time $t$ was $\lambda_{\rm max}(t)\sim 2\pi R_{\rm h}(t)$. In the $R_{\rm h}=ct$ Universe,
the comoving distance to the last scattering surface (at time $t_e$) is
\begin{equation}
 r_e=ct_0\int_{t_e}^{t_0}{dt'\over t'}=ct_0\ln\left({t_0\over t_e}\right)\;.
\end{equation}
Therefore, the maximum angular size $\theta_{\rm max}$ of any fluctuation associated
with the CMB emitted at $t_e$ has to be
\begin{equation}
\theta_{\rm max}={\lambda_{\rm max}(t_e)\over R_e(t_e)}\;,
\end{equation}
where
\begin{equation}
R_e(t_e)=a(t_e)r_e=a(t_e)ct_0\ln\left({t_0\over t_e}\right)=ct_e\ln\left({t_0\over t_e}\right)
\end{equation}
is the proper distance to the last scattering surface at time $t_e$. That is,
\begin{equation}
\theta_{\rm max}\sim {2\pi\over \ln(t_0/t_e)}\;.
\end{equation}
Thus, if we naively adopt the times $t_0=13.7$ Gyr and $t_e\approx 380,000$ yrs from
the standard model, we find that $\theta_{\rm max}\sim 34^\circ$. As we shall see, it
is the existence of this limit, more than any other aspect of the CMB anisotropies in the $R_{\rm h}=ct$
Universe, that accounts for the shape of the observed angular correlation function
in figure~1.

Since we are here beginning to identify specific values for the age of the
Universe $t_0$, and the recombination time $t_e$, both of which impact observables, 
such as $\theta_{\rm max}$, it is important to remind ourselves that the concordance 
values of the $\Lambda$CDM parameters render the expansion history in the standard model
so similar to that in $R_{\rm h}=ct$ that today we measure $R_{\rm h}(t_0)\approx ct_0$. 
In other words, the age of the concordance $\Lambda$CDM Universe is virtually identical 
to that of the $R_{\rm h}=ct$ Universe, so using $t_0=13.7$ Gyr for these estimates is quite reasonable. 

Insofar as the radiation in $R_{\rm h}=ct$ is concerned, its temperature increases 
inversely with $a(t)$, as it does in $\Lambda$CDM, so there is little difference 
in the radiation fields within these two cosmologies. What does differ is the dark-energy
content and its equation of state. In $\Lambda$CDM, one typically makes the simplest
assumption, which is that dark energy is a cosmological constant and therefore
becomes less important as $t\rightarrow 0$. In $R_{\rm h}=ct$, on the other hand,
all that matters is that the constituents together must contribute an overall equation 
of state $p=-\rho/3$, so any components other than matter and radiation have a stronger
dependence on $a(t)$ (and therefore $t$) than they do in $\Lambda$CDM. However, the 
radiation will still have the same temperature at a given value of $a(t)$
as it does in $\Lambda$CDM. This suggests that the range over which $t_e$ will fall
in $R_{\rm h}=ct$ is probably not far from $380,000$ yrs. Below, we will consider
the impact on $C(\cos\theta)$ from changes to the ratio $t_0/t_e$ over the range
$5\times 10^4-5\times 10^6$ (see figure~4).

In this paper, we wish to identify the key elements of the theory responsible for the
shape of $C(\cos\theta)$, without necessarily getting lost in the details of the complex treatment
alluded to in \S3 above, so we will take a simplified approach used quite effectively in
other applications (Efstathiou 1990). For example, we will ignore the transfer
function, and consider only the Sachs-Wolfe effect, since previous work has shown that
this is dominant on scales larger than $\sim 1^\circ$. From a practical standpoint, this
approximation affects the shape of $C(\cos\theta)$ closest to $\theta=0$ in figure~1,
but that's not where the most interesting comparison with the data will be made. We will
also adopt a heuristic, Newtonian argument to establish the scale-dependence of this
effect, noting that
\begin{equation}
\Delta\Phi\sim {G\Delta M\over \lambda}\;,
\end{equation}
where
\begin{equation}
\Delta M={4\pi\over 3}\delta\rho\lambda^3\;.
\end{equation}
Thus, from Eq.~(7), we see that
\begin{equation}
{\Delta T\over T}\sim \delta\rho\lambda^2\;.
\end{equation}
But the variance in density over a particular comoving scale $\lambda$ is given as
\begin{equation}
\left({\delta\rho\over \rho}\right)^2_\lambda\propto
\int_0^{\kappa\sim1/\lambda}P(\kappa')d^3\kappa'
\end{equation}
(Efstathiou 1990). Not knowing the exact form of the power spectrum
emerging from the non-linear growth prior to recombination, we will parametrize it
as follows,
\begin{equation}
P(\kappa)\propto \kappa-b\left({2\pi\over R_e(t_e)}\right)^2\kappa^{-1}\;,
\end{equation}
where the (unknown) constant $b$ is expected to be $\sim O(1)$. 

   \begin{figure}[hp]
   \centering
      \includegraphics[angle=0,width=8.2cm]{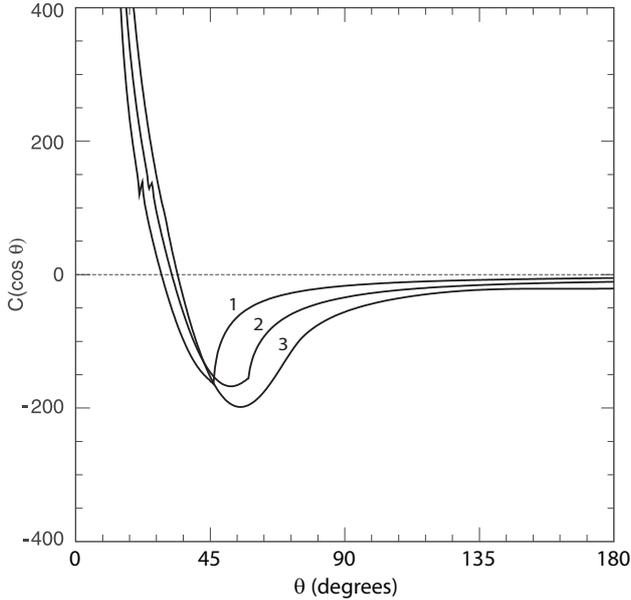}
      \caption{
Impact on the angular correlation function $C(\cos\theta)$ in the $R_{\rm h}=ct$
Universe from a change in the ratio $t_0/t_e$: (Curve 1), $5\times 10^6$; (Curve 2),
$5\times 10^5$; (Curve 3), $5\times 10^4$.}
\end{figure}

This form of $P(\kappa)$ is based on the following reasoning.
The conventional procedure is to assume a scale-free initial 
power-law spectrum, which we also do here. In $\Lambda$CDM, these
fluctuations grow and then expand to all scales during the required
inflationary phase. In $R_{\rm h}=ct$, the fluctuation growth is driven 
by the (negative) pressure, represented by the term on the right-hand
side of Eq.~(10). Very importantly, because there is no Jeans length,
fluctuations can in principle grow on all scales. However, this equation
also shows that what matters most is the ratio of the fluctuation length
$\lambda$ to the gravitational radius $R_{\rm h}(t)$ at time $t$. 
The solution to this equation shows that only fluctuations with $\lambda 
< 2\pi R_{\rm h}$ will grow, and that those modes that grow, will grow rapidly, 
given their strong dependence on $t$ (see Eq. 16). 

The most important result we get from this analysis is that fluctuations
will grow quickly in amplitude until they get to the size $2\pi R_{\rm h}(t)$,
and then the growth stops. In $\Lambda$CDM, on the other hand, growth continues 
due to inflation. The simple parametrization in Eq.~(27) incorporates these 
essential effects: first, the initial seed spectrum is assumed to be
scale-free, which means that $P(\kappa)\sim\kappa$. Since the growth rate 
depends critically on the ratio $R_{\rm h}/\lambda$, one would expect 
$P(\kappa)$ to be dominated by the smaller wavelengths (i.e., the
larger $\kappa$'s), and be altered more and more for increasing
wavelengths (i.e., smallter $\kappa$'s). Thus, for large $\kappa$, 
$P(\kappa)$ should still go roughly as $\kappa$. However, for smaller 
$\kappa$, the growth rate decreases with decreasing $\kappa$, so one 
would expect a greater and greater depletion in power with decreasing 
$\kappa$. The second term in Eq.~(27) represents this effect.

Having said this, the key physical element most responsible for the shape 
of the angular correlation function is the ratio $t_0/t_e$, since this
determines the size of the gravitational horizon $R_{\rm h}(t_e)$ from 
which one determines the maximum angle $\theta_{\rm max}$ of the 
fluctuations. The $b$ in the parametrization of Eq.~(27) affects the 
location of the minimum in (and value of) $C(\cos\theta)$, but not 
qualitatively, as exhibited by the variations shown in figure~5. Since
the results are only weakly dependent on $b$, the parametrization
in Eq.~(27) does not appear to be overly influencing our results.

It is not difficult to see
from Eqs.~(26) and (27) that $\delta\rho$ is therefore given as
\begin{equation}
\delta\rho\sim {1\over\lambda^2}\left(1-b\theta^2\right)\;,
\end{equation}
where the definition of $\theta$ is analogous to that of $\theta_{\rm max}$ in
Eq.~(20). Thus, the amplitude of the Sachs-Wolfe temperature fluctuations
follows the very simple form
\begin{equation}
{\Delta T\over T}\sim \left(1-b\theta^2\right)\;,
\end{equation}
but only up to the maximum angle $\theta_{\rm max}$ established earlier. The CMB angular correlation
function $C(\cos\theta)$ calculated using Eq.~(29) is shown in figure~2, together with the
{\it Planck} data (Planck Collaboration XV 2013), for $b=3$ and $t_0/t_e=5\times 10^5$. 
To facilitate a comparison between all three correlation functions (from {\it Planck}, 
$\Lambda$CDM, and $R_{\rm h}=ct$), we show them side by side in figure~3.

We are not yet in a position to calculate the probability of getting the 
observed $S_{1/2}$ for all possible realizations of the $R_{\rm h}=ct$ Universe, 
because our estimation of the fluctuation spectrum in this cosmology is still at 
a very rudimentary stage. Our calculation of the correlation function is based
solely on the Sachs-Wolfe effect, which dominates at large angles. However,
the general agreement between theory and observation in figures~2 and 3
suggests that the angular correlation function associated with an FRW
cosmology without inflation (such as $R_{\rm h}=ct$) matches that observed
with WMAP and {\it Planck} qualitatively better than $\Lambda$CDM. For
example, we note from figure~3 that $R_{\rm h}=ct$ does a better job
predicting the location of the minimum, $\theta_{\rm min}$, the value of 
$C(\cos\theta_{\rm min})$ at this angle and, particularly, the lack of
significant angular correlation at angles $\theta>60^\circ$.

Of course, one should wonder how sensitively
any of these results depends on the chosen values of $b$ and $t_0/t_e$. The short answer is
that the dependence is weak, in part because the ratio $t_0/t_e$ enters into the calculation only
via its log (see  Eq.~22). Figure~4 illustrates the impact on $C(\cos\theta)$ of changing
the value of $t_0/t_e$, from $5\times 10^6$ (curve 1), to $5\times 10^5$ (curve 2),
and finally to $5\times 10^4$ (curve 3).

   \begin{figure}[h]
   \centering
      \includegraphics[angle=0,width=8.2cm]{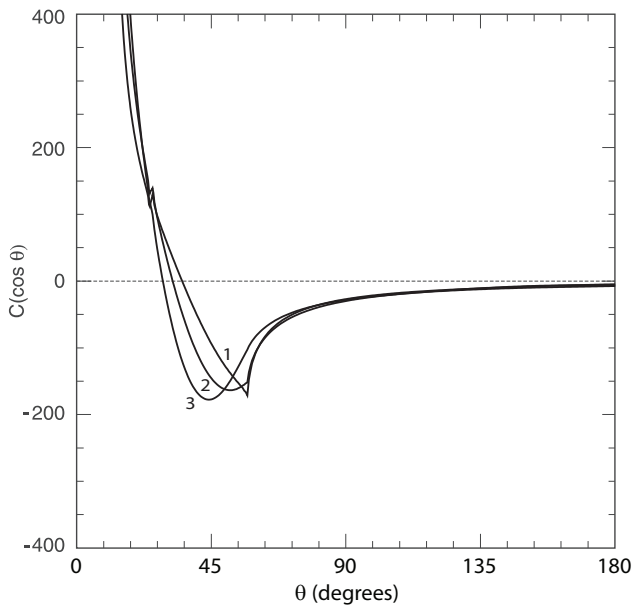}
      \caption{
Impact on the angular correlation function $C(\cos\theta)$ in the $R_{\rm h}=ct$
Universe from a change in the value of $b$: (Curve 1), $b=2$; (Curve 2),
$b=3$; (Curve 3), $b=4$.}
\end{figure}

The dependence of these results on the value of $b$ is illustrated in figure~5, which shows
the curves corresponding to $b=2$ (curve 1), $b=3$ (curve 2), and $b=4$ (curve 3). In all
cases, the overarching influence is clearly the existence of $\theta_{\rm max}$, which cuts
off any angular correlation at angles greater than $\sim 60^\circ$. We would argue that any
of the curves in Figs.~3 and 4 present a better match to the WMAP data than the
$\Lambda$CDM curve shown in figure~1. Therefore, the difference between $\Lambda$CDM
and the data may be due to inflation rather than cosmic variance.

Finally, let us acknowledge the fact that although $\Lambda$CDM may not
appear to provide the better explanation for the angular correlation function,
it nonetheless does extremely well in accounting for the observed angular
power spectrum for $l>20$ (see, e.g., Bennett et al. 2013; Planck Collaboration
XV 2013). Along with its remarkable fit to the Type Ia SN data, this
has been arguably the biggest success story of this long-standing 
cosmological model. We have not yet included all of the physical effects,
such as baryon acoustic oscillations, occurring near the surface of last scattering
in the $R_{\rm h}=ct$ scenario, so it is not yet possible to carry out a complete
comparative analysis of the entire power spectrum between the two models,
certainly not for $l>10-20$. The work of Scott et al. (1995), among others, 
suggests that, unlike the Sachs-Wolfe effect, which is quite sensitive to the 
expansion dynamics, the local physics where the CMB is produced may be 
generic to a wide range of evolutionary histories. So the fact that 
$\Lambda$CDM does not acccount very well for the angular correlation 
function, which tends to highlight features predominantly at large angles
($\theta>1-10$ degrees), is not inconsistent with the reality that it
fits the high-$l$ angular power spectrum very well.

To demonstrate how these two approaches to the analysis of angular
information in the CMB focus on quite different aspects of the
fluctuation problem, we show in figure~6 the angular power spectrum
produced solely by the Sachs-Wolfe effect in the $R_{\rm h}=ct$
Universe.  This fit for $l<10-20$ is actually a better match to the observations 
than that associated with $\Lambda$CDM, but what is lacking, of course,
is information for $l>10-20$, where the standard model does exceptionally
well. It is comforting to see that the qualitatively good fit exhibited by
$R_{\rm h}=ct$ in figures~2 and 3, is confirmed by the very good fit
also seen in the angular power spectrum (figure~6) at low values of $l$.
(The details of how this angular power spectrum is calculated are provided
in a companion paper, whose principal goal is to discuss the apparent
low-multipole alignment in the CMB; see Melia 2012.)
Future work must include a more complete analysis than we have 
presented here, to approach the extraordinary level of detail now
available in applications of the standard model.

   \begin{figure}[h]
   \centering
      \includegraphics[angle=0,width=8.2cm]{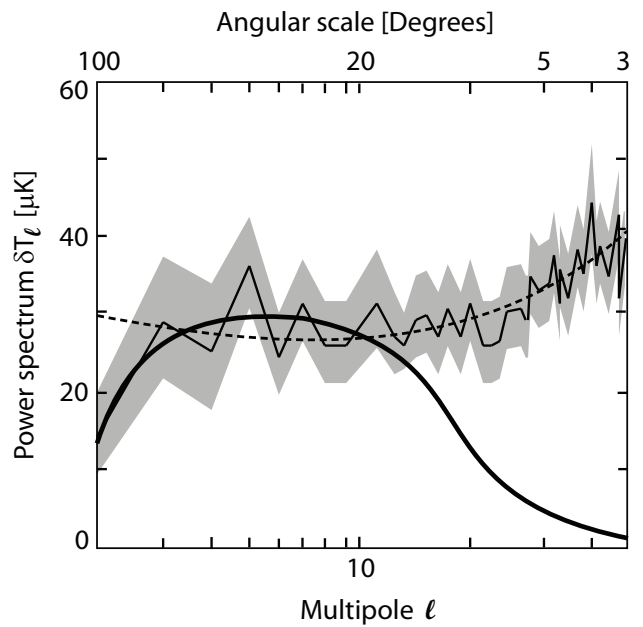}
      \caption{
     The theoretical CMB power spectrum due solely to
      Sachs-Wolfe-induced fluctuations in the $R_{\rm h }=ct$
      Universe (solid, thick curve), in comparison with the
      power spectrum measured from the full WMAP sky
      (thin, jagged line; Spergel et al. 2003; Tegmark et al. 2003).
      The gray region represents the one-sigma uncertainty. 
      The power spectrum for $l> 10-20$ is dominated by 
      small-scale physical effects, such as baryon acoustic 
      oscillations near the surface of last scattering, which are not 
      included in our analysis. For comparison, we also show 
      the WMAP best-fit (dashed) curve, calculated with all of the
      physical effects producing the fluctuations. This curve
      fits the data very well, particularly at very high $l$'s, 
      corresponding to fluctuations on scales smaller than 
      $\sim 10^\circ$.}
\end{figure}

\section{Conclusions}
It is essential for us to identify the key physical ingredient that guides the behavior
of a diagnostic as complicated as $C(\cos\theta)$ in Figs.~1 through 5. An episode of
inflation early in the Universe's history would have driven all fluctuations to grow,
whether in the radiation dominated era, or later during the matter dominated expansion,
to very large opening angles, producing a significant angular correlation on all
scales. The {\it Planck} data reproduced in figure~1 (and also the earlier WMAP 
observations) show that this excessive expansion  may not have occurred, if 
the difference between theory and observations is not
due solely to cosmic variance. In the $R_h=ct$ Universe,
on the other hand, there was never any inflationary expansion (Melia 2013a),
so there was a limit to how large the fluctuations could have grown by the time ($t_e$)
the CMB was produced at the surface of last scattering. This limit was attained
when fluctuations of size $\lambda/2\pi$ had reached the gravitational horizon
$R_{\rm h}(t_e)$. And for a ratio $t_0/t_e\sim  35,000$--$40,000$, this
corresponds to a maximum fluctuation angle $\theta_{\rm max}\sim 
30$--$35^\circ$. This limit is the key ingredient responsible for the shape of the
angular correlation function seen in Figs.~2, 3, 4 and 5. Though other
influences, such as Doppler shifts, the growth of adiabatic perturbations, and the
integrated Sachs-Wolfe effect, have yet to be included in these calculations,
they are not expected---on the basis of previous work---to be dominant; they
would modify the shape of $C(\cos\theta)$ only slightly (perhaps even bringing the
theoretical curve closer to the data). The positive comparison between the 
observed and calculated $C(\cos\theta)$ curves seen in these figures offers some
support to the viability of the $R_{\rm h}=ct$ Universe as the correct description 
of nature.

One might also wonder whether the observed lack of angular correlation and the
alignment of quadrupole and octopole moments are somehow related. This question
was the subject of Raki\'c \& Schwarz's analysis (Raki\'c \& Schwarz 2007), which concluded that 
the answer is probably no. More specifically, they inferred that having one does
not imply a larger or smaller probability of having the other. However, this analysis
was rather simplistic, in the sense that it did not consider whether alternative
cosmologies, such as the $R_{\rm h}=ct$ Universe, could produce the observed
alignment as a result of the $\sim R_{\rm h}(t_e)$ (non-inflated) fluctuation-size
limit, in which case the two anomalies would in fact be related, though only
indirectly. 

In related work, it was shown by Sarkar et al. (2011) that there is no statistically 
significant correlation in $\Lambda$CDM between the missing power on large angular 
scales and the alignment of the $l=2$ and $l=3$ multipoles. If not due to variance,
the inconsistency between the standard model and the WMAP data may therefore be
greater than each of the anomalies alone, because their combined statistical 
significance is equal to the product of their individual significances. As pointed 
out by Sarkar et al. (2011), such an outcome would require a causal explanation.

In this paper, we have shown that at least one of these anomalies is not generic
to all FRW cosmologies. In fact, the observed angular correlation function is
a good match to that predicted in the $R_{\rm  h}=ct$ Universe. This property of 
the CMB might be pointing to the existence of a maximum angular size 
$\theta_{\rm max}$ for the large-scale fluctuations, imposed by the 
gravitational horizon $R_{\rm h}$ at the time $t_e$ of last scattering.

\begin{acknowledgements}
This research was partially supported by
ONR grant N00014-09-C-0032 at the University of Arizona, and by a Miegunyah
Fellowship at the University of Melbourne. I am particularly grateful to
Amherst College for its support through a John Woodruff Simpson Lectureship.
And I am happy to acknowledge the helpful comments by the anonymous 
referee, that have led to a significant improvement in this manuscript.
\end{acknowledgements}


\begin{thebibliography}{}

\bibitem[]{Bennett03}
Bennett, C. L. et al. 2003, ApJS, 148, 97

\bibitem[]{Bennett13}
Bennett, C. L. et al. 2013, ApJS, in press (arXiv:1212.5225)

\bibitem[]{Copi09}
Copi, C. J., Huterer, D., Schwarz, D. J. \& Starkman, G. D.
2009, MNRAS, 399, 295

\bibitem[]{Copi13}
Copi, C. J., Huterer, D., Schwarz, D. J. \& Starkman, G. D.
2013, MNRAS, in press (arXiv:1310.3831)

\bibitem[]{Efstathiou90}
Efstathiou, G. 1990, in Physics of the Early Universe, eds. J. A. Peacock,
A. F. Heavens \& A. T. Davies, SUSSP, Edinburgh.

\bibitem[]{Guth81}
Guth, A. H. 1981, Phys. Rev. D, 23, 347

\bibitem[]{Hinshaw96}
Hinshaw, G. et al. 1996, ApJ Letters, 464, L25

\bibitem[]{Linde82}
Linde, A. 1982, Phys. Lett. B, 108, 389

\bibitem[]{Melia07}
Melia, F. 2007, MNRAS, 382, 1917

\bibitem[]{Melia12}
Melia, F. 2012, ApJ, submitted (arXiv:1207.0734)

\bibitem[]{Melia13a}
Melia, F. 2013a, A\&A, 553, id A76

\bibitem[]{Melia13b}
Melia, F. 2013b, Phys. Lett. B, submitted

\bibitem[]{Melia13c}
Melia, F. 2013c, CQG, 30, 155007

\bibitem[]{Melia09}
Melia, F. and Abdelqader, M. 2009, IJMP-D, 18, 1889

\bibitem[]{MeliaMaier13}
Melia, F. \& Maier, R. S. 2013, MNRAS, 432, 2669 

\bibitem[]{MeliaShevchuk12}
Melia, F. \& Shevchuk, A. 2012, MNRAS, 419, 2579

\bibitem[]{Moresco12}
Moresco, M., Verde, L., Pozzetti, L., Jimenez, R. \& Cimatti, A. 2012,
JCAP, in press (arXiv:1201.6658)

\bibitem[]{Nolta09}
Nolta, M. R. et al. 2009, ApJS, 180, 296

\bibitem[]{Planck13}
Planck Collaboration XV, 2013, A\&A submitted (arXiv:1303.5075)

\bibitem[]{Rakic07}
Raki\'c, A. \& Schwarz, D. J. 2007, Phys. Rev. D, 75, Article ID 103002

\bibitem[]{Rees68}
Rees, M. J. \& Sciama, D. W. 1968, Nature, 217, 511

\bibitem[]{Sachs67}
Sachs, R. K. \& Wolfe, A. M. 1967, ApJ, 147, 73

\bibitem[]{Sarkar11}
Sarkar, D., Huterer, D., Copi, C. J., Starkman, G. D. \&
Schwarz, D. J. 2011, Astroparticle Phys., 34, 591

\bibitem[]{Scott95}
Scott, D., Silk, J. \& White, M. 1995, Science, 268, 829

\bibitem[]{Spergel03}
Spergel, D. N. et al. 2003, ApJS, 148, 175

\bibitem[]{Sunyaev80}
Sunyaev, R. A. \& Zeldovich, I. B. 1980, MNRAS, 190, 413

\bibitem[] {Tegmark03}
Tegmark, M., de Oliveira-Costa, A. \& Hamilton, A. J., 2003, Phys. Rev. D, 68, Article ID 123523

\bibitem[]{Wright96}
Wright, E. L., Bennett, C. L., Gorski, K., Hinshaw, G. \& Smoot, G. F.
1996, ApJ Letters, 464, L21

\bibitem[]{Watson11}
Watson, D. F., Berlind, A. A. \& Zentner, A. R. 2011, ApJ, 738, article id. 22

\bibitem[]{Wei13}
Wei, J.-J., Wu, X. \& Melia, F. 2013, ApJ, 772, 43

\end{thebibliography}
\end{document}